\documentstyle[twocolumn,aps]{revtex}
\begin{document}
\draft
\twocolumn[\hsize\textwidth\columnwidth\hsize\csname@twocolumnfalse\endcsname
\title{Separatrix Reconnections in Chaotic Regimes}
\author{G. Corso\\
Department of Electrical Engineering and \\
Computer Science, and the Electronics  
Research Laboratory,\\
University of California, Berkeley, California 94720-1770\\ 
and\\ F.B. Rizzato\\
Instituto de F\'{\i}sica Universidade Federal do Rio Grande do Sul\\
P.O.Box 15051 91501-970 Porto Alegre, RS, Brazil}
\maketitle
\date{\today}
\begin{abstract}
In this paper we extend the concept of separatrix reconnection into 
chaotic regimes. We show that even under chaotic conditions one can still 
understand abrupt jumps of diffusive-like processes in the relevant phase-space 
in terms of relatively smooth realignments of stable and unstable manifolds 
of unstable fixed points.
\end{abstract}
%
]
%
Two dimensional nonmonotonic conservative maps are recognized to be of 
relevance in modeling a number of nonlinear systems, as for instance laser 
acceleration of charged particles \cite{how84,car92,ita92,how95,nos95,lili91}, 
and the 
nonlinear flow of magnetic field lines in fusion machines like tokamaks and 
others \cite{oda96,cor97}. As opposed to the more traditional monotonic versions, 
nonmonotonic maps are characterized by frequency curves that are not monotonic 
functions of the action variable. In laser accelerators, nonmonotonicity 
arises as a result of the relativistic mass variation of the accelerating 
particles \cite{nos95}; in tokamaks, it arises as a result of the geometrical 
peculiarities of the relevant background magnetic fields. In any case, 
nonmonotonicity has strong influence on the types of bifurcations that can occur 
in the associated nonlinear dynamics. Indeed, monotonic maps typically allow 
only period doubling bifurcations of fixed points but if one adds nonmonotonicity, 
tangent bifurcations involving pairs of elliptic and hyperbolic fixed points 
also become possible in the appropriate phase-space. 

\begin{figure}[h]
\caption{Inverse tangent bifurcation and the preceding reconnection in a purely 
integrable case. The figure is constructed with an integrable 
one-degree-of-freedom Hamiltonian of the type $H = p - \alpha p^2 + k \cos \phi$ 
with $(p,\phi)$ as continuous canonical variables; $k=5$ and 
$\alpha=0.10$ in (a), $0.128$ in (b), and $0.16$ in (c).}
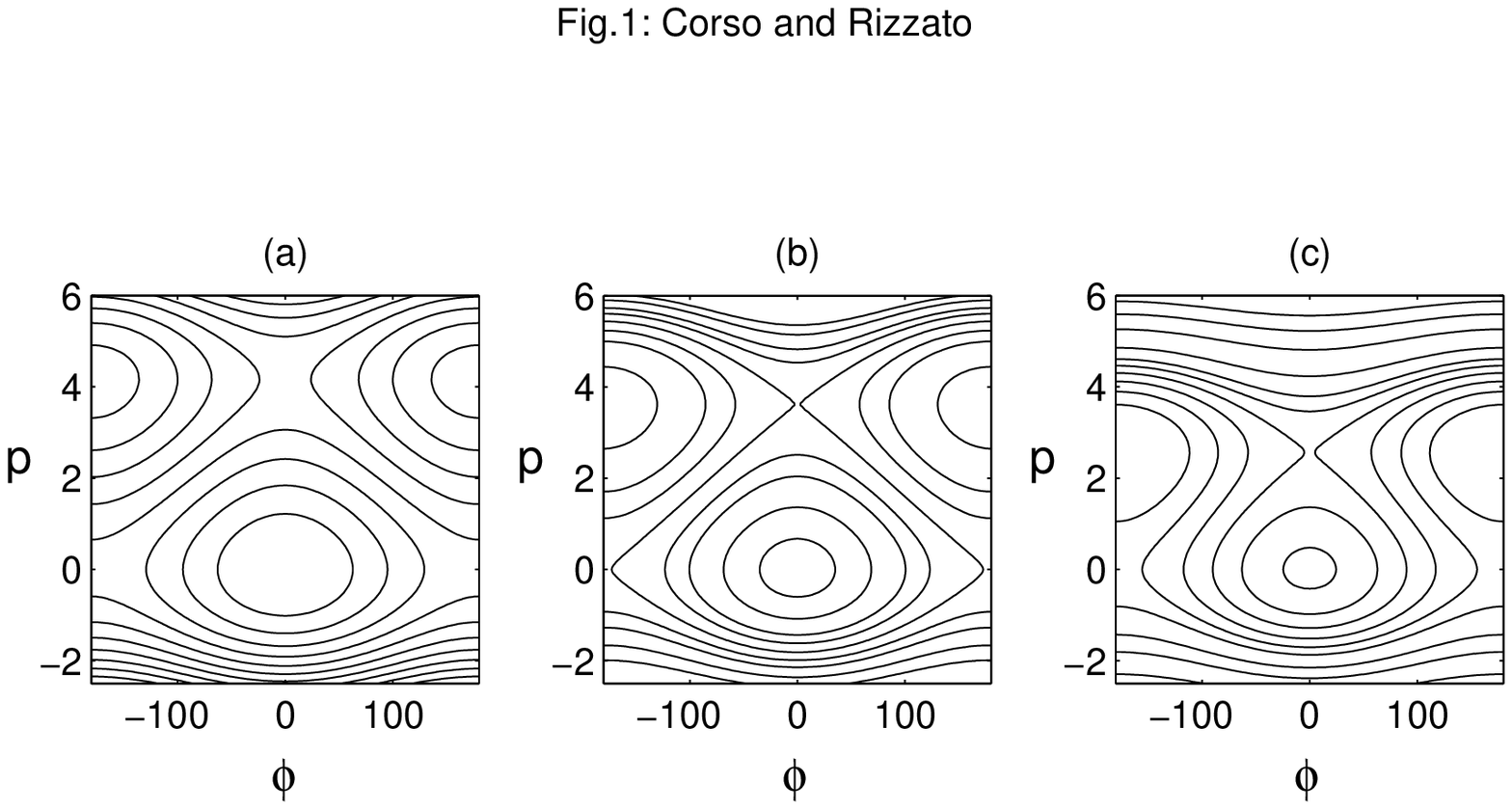
\label{fig1.ps}
\end{figure}

Period doubling cascades of periodic orbits generally precede a transition 
to chaotic regimes of these orbits, but tangent bifurcations bear no 
direct relationship to nonintegrability. Indeed, it has been argued that the 
most noticeable effects of a tangent bifurcation are to be seen while the 
system lies in predominantly integrable regimes. Although we will show that 
this sort of argument is questionable, consider the process depicted in the 
integrable case of Fig.(\ref{fig1.ps}) \cite{car92}. 
Two chains of fixed points undergo tangent 
bifurcation, or actually an inverse tangent bifurcation, starting from the 
leftmost panel down to the rightmost. Before the bifurcation itself where 
elliptic fixed points collapse against hyperbolic points, the separatrices 
defining the upper chain undergo a reconnecting process with those defining the 
lower chain - this is seen in Fig.(\ref{fig1.ps}b). 
It is precisely due to the smoothness caused by 
integrability that the reconnection can be seen so clearly. 
This is why reconnection is thought to be of relevance only in integrable cases. 
In contrast, the process is generally regarded as of little significance in 
chaotic regimes because in those situations all separatrices - which shall be 
correctly called stable and unstable manifolds then - would be already 
interlaced with little global response as relevant control parameters are varied. 
Speaking in more precise terms, effects associated with reconnections are thought 
to be unobservable when the elliptic fixed points of the reconnecting chains 
undergo full cascades of period doublings, {\it before} any sort of mutual 
contact of the relevant manifolds takes place. 

While it is true that reconnections are not easily visualized in chaotic regimes, 
it is our purpose here to show that their effect can still be quite appreciable. 
What happens is that even in chaotic regimes the unstable or stable manifolds of 
originally hyperbolic points may still make a transition from a situation where 
their mutual crossings are absent - or, actually, relatively infrequent - to a 
situation where the mutual crossings become very frequent. We shall illustrate 
the process with a set of figures from which one will be able to see that this 
change in topology is a reminiscent of the corresponding behavior of regular 
regimes. The macroscopic result of this type of transition is that as soon as 
the crossings become frequent, stochastic diffusion undergoes an abrupt jump 
from slower to faster rates. 

From this point on, the discussion relies on more technical grounds. 
Therefore we introduce here the model map we shall be working with. The map is 
called the nonmonotonic twist map and reads: 
\begin{equation}
p_{n+1} = p_n - k \sin \phi_n \>,\>\>\> \phi_{n+1} = \phi_n + f(p_{n+1}),
\label{mapa}
\end{equation}
where $(p,\phi)$ is a pair of discrete canonical variables - $p$ representing an 
action and $\phi$ a $2 \pi$-periodic angular coordinate - and where 
the map itself is totally simpletic given that the left-hand-side of 
the second Eq.(\ref{mapa}) depends on $p_{n+1}$ instead of $p_n$. Function 
$f(p)$ is of foremost importance here. It is in fact a measure of the frequency 
with which the discrete orbits move on the $(p,\phi)$ phase-space. In standard 
monotonic maps it reads $f(p)=p$, but as we wish to incorporate nonmonotonic 
features we add a quadratic term such that $f$ becomes $f(p) = p - \alpha p^2$ as 
in Ref.\cite{how84}. In this case one has effectively a nonmonotonic frequency 
curve with maximum located at $p=1/(2 \alpha)$, where $df(p)/dp=0$. The map 
(\ref{mapa}) has several families of fixed points. Let us focus here on the 
first order family (period one orbits) which is characterized by 
$p_{n+1} = p_n \>\>\> {\rm and} \>\>\> \phi_{n+1} = \phi_n + 2\,m\,\pi$,
with $m$ as an arbitrary positive or negative integer; we shall refer to 
the fixed points as $(p_m^\ast,\phi_m^\ast)$.
Eq.(\ref{mapa}) informs that the fixed points are located at 
$\phi_m^\ast=0,\pi \>\>\> {\rm and} \>\>\> p_m^\ast = 
[1 \pm \sqrt{1 - 8 m \pi \alpha}/(2 \alpha)]$.
In this case of period one orbits, let us have a brief look at 
the distribution of the various fixed points over the phase-space. 
For $m=0$ one has four points located at $p_o^\ast = 0,\,1/\alpha$. 
If $m>0$ the fixed points lie in the finite interval $0 < p < 1/\alpha$, 
and if $m<0$ the points lie in any of the intervals $-\infty < p < 0$ or 
$1/\alpha < p < +\infty$. The existence of points located in the finite interval 
must satisfy the condition $m < 1/(8 \pi \alpha)$, so it may well 
happen that no fixed point can be actually found there if $\alpha$ is large 
enough. What happens is that as $\alpha$ grows from some small value, all the 
fixed points originally located in the sub-interval $0<p<1/(2\alpha$), collapse, 
at $p=1/(2\alpha)$, against the corresponding points originally located in the 
sub-interval $1/(2\alpha)<p<1/\alpha$ via a sequence of inverse tangent 
bifurcations. Meanwhile, the points placed externally to the interval 
$(0,1/\alpha)$, as well as the points of the pair $m=0$, simply approach each 
other but never touch. We point out that although this latter two types of fixed 
points never undergo inverse tangent bifurcation, their manifolds can 
naturally undergo the reconnection processes. In previous works \cite{how84} the 
$m=1$ case has been investigated. It has been shown that when chaos is absent, 
in the sense that elliptic points of the various chains have not yet period 
doubled to chaos, initial conditions at negative values of $p$ do not 
move up to positive values unless the separatrices of the two chains 
corresponding to the $m=1$ resonances touch each other. In addition it has been 
argued that for those situations where the elliptic fixed points have already 
undergone full cascades of period doublings, no detectable difference in the 
global aspects of the dynamics should be observed as arising from a possible 
contact of separatrices. We now proceed to show that this is not quite exact; 
even under chaotic conditions, some noticeable effects resulting from manifold 
reconnections can be in fact observed. Specifically we shall show that after 
what could be best called a reconnection-like process, diffusion makes a 
somewhat abrupt jump from lower to higher values. 

Let us focus the discussion on the $m=0$ case because this resonance is 
the largest one in the system. The linear stability of the fixed points 
can be examined from the characteristic equation
$\lambda = [(2-\kappa) \pm \sqrt{(\kappa-2)^2 -4}] / 2$,
with $\kappa \equiv \cos(\phi_o^\ast) (1-2 \alpha p_o^\ast) k$.
$\lambda$ is the eigenvalue of the linearized map; if complex (purely real) the 
corresponding fixed point is unstable (stable). One then sees 
that for the chain located at $p_o^\ast = 0$, the fixed point at 
$\phi_o^\ast = \pi$ is always unstable, while that at $\phi_o^\ast = 0$ is 
unstable only when $k > 4$, being stable otherwise. As for the chain at 
$p_o^\ast = 1/\alpha$, the point at $\phi_o^\ast=0$ 
is always unstable, while that at $\phi_o^\ast= \pi$ is unstable when $k > 4$; 
in both cases, de-stabilization occurs via period doubling of the elliptic 
points. One can also make an estimate of the condition to be observed for 
separatrix touching. To do so, let us first imagine that we are working in a 
situation where $k$ is small and the dynamics is therefore mostly regular. What 
happens then is that the advance within any particular resonant island tends to 
be slow. One can thus approximate $p_{n+1}-p_n$ and $\phi_{n+1}-\phi_n$ by 
their respective infinitesimal increments $dp$ and $d\phi$ and finally write 
an expression valid in the vicinity of a $m=0$ chain:
$dp / d\phi = -k \sin(\phi) / f(p)$.
Then, with obvious notation, one obtains an expression for the separatrix of 
the lower chain:
\begin{equation}
\int^{p_{sep}} f(p) dp = k (1 + \cos(\phi_{sep})).
\label{sepint}  
\end{equation}

The upper separatrix of the ($p_{o}=0,\phi_{o}=\pi$) fixed point touches 
the ($p_{o}=1/\alpha,\phi_{o}=0$) fixed point of the corresponding upper 
chain (see Fig.(\ref{fig1.ps})) when 
$k=k_t \equiv 1/(12 \alpha^2)$; such value for $k$ 
is also known as the reconnection threshold in regular regimes. We shall extend 
this definition for the threshold into chaotic regimes. If we do that we can 
draw Fig.(\ref{fig2.ps}). The figure is similar to corresponding 
figures shown in Refs. \cite{how84,how95}. It displays simultaneously the 
threshold and period doubling curves in the parameter space. The period 
doubling curve, in the $m=0$ case analyzed here, is simply a horizontal line 
at $k=4$. It is thus seen that for values of $\alpha$ below 
$\alpha \sim 0.144$ period doublings occur before reconnection. Previous works 
have focused interest on the region $\alpha > 0.144$ because in that region 
one could clearly speak in terms of reconnection - recall that in this region 
reconnection takes place before period doublings. In the present 
paper we shall concentrate efforts to see what happens deep into chaotic 
regimes when $\alpha < 0.144$.

To start with the investigation, let us consider the vicinity of the threshold 
curve at $k = 5$. For this value of $k$, elliptic points have been totally 
destroyed by full cascades of period doublings. 
Therefore, as we increase $\alpha$ the theoretical reconnection threshold 
can be attained while the system remains in a deep chaotic 
regime. Let us try to examine how the relevant manifolds of the 
$m=0$ resonances behave on the phase-space. The analysis is made with help of 
the panels of Fig.(\ref{fig3.ps}) were we focus attention on the upper unstable 
and lower stable manifolds (respective orientation indicated by arrows) of the 
originally hyperbolic points (points indicated by black dots) of the lower and 
upper chain respectively; in order to draw the manifolds we launch 1000 initial 
conditions along the linearized manifolds, iterating the dynamics forward or 
backwards according to the case.

First of all we note that in integrable cases separatrices describe homoclinic 
loops. Now, even in our nonintegrable case, when $\alpha$ is small it is seen 
that the tendency of the unstable manifold of the lower chain is to follow the 
homoclinic loop of the integrable approximation. Of course, due to 
the nonintegrable features the unstable manifold eventually starts to execute 
increasingly large oscillations after it first intersects the stable manifold of 
the same point. As the orbit is about to complete the homoclinic loop the 
oscillations grow and as the oscillations grow it may happen that the unstable 
manifold of the lower chain crosses the stable manifold of the upper chain as 
well. But what must be observed here is that this latter intersection occurs only 
after the stable manifold of the lower chain crosses the unstable manifold of 
the same lower chain many times. In this situation one can safely look at 
the process as closely resembling the integrable case, although, 
as mentioned before, one lies in a deeply chaotic regime; this feature is 
somewhat puzzling and shall be considered in detail later.

Then, when one increases the value of $\alpha$, the overall topology of 
manifold crossings appears to undergo a substantial change. This can be 
observed in Fig.(\ref{fig3.ps}b) where it is seen that this change in topology 
is in fact very similar to what happens in the purely 
integrable model; to our knowledge this had not been realized before. Here 
the unstable manifold of the hyperbolic point of the lower chain 
makes direct connection with the stable manifold of the upper chain. This 
leads to the opening of a new diffusive channel connecting the regions located 
below the lower chain and above the upper chain. In other words, although 
one lies in a deep chaotic regime since the original elliptic points 
have fully period doubled to chaos, noticeable changes can be 
expected as a result of clear alterations on the topology of the 
manifolds. This behavior of the manifolds is rather conservative in the 
sense that while the elliptic points have bifurcated, the manifolds still 
try to preserve some aspects of integrability. As it appears this feature takes 
place because near the midpoint between the two chains, the orbital frequency 
attains a maximum $d f(p) / d p = 0$ in view of nonmonotonicity. As a result, 
the local dynamics is relatively linear and nonintegrable effects are 
relatively smaller than in other regions of the phase-space.

We now proceed to show that reconnections in chaotic regimes may have 
direct influence on macroscopic processes, such as diffusion. We actually 
measure a fraction of particles that are transmitted across the region where 
$d f (p) / d p \sim 0$ in a numerical experiment that goes as follows.

A set of 1000 initial conditions is launched in the region below 
the lower reconnecting chain and iterated many times. All particles arriving 
at the region above the upper reconnecting chain are reinjected into the 
initial lower region such as to create a steady state in the long run. The 
simulation is optimized by reflecting particles that move into the 
region below the injection momentum $p=-2.0$. The mental picture one can form of 
the system is that of a multicomponent gas placed in a vessel divided by a 
semi-permeable membrane whose role is played by the reconnecting chains. 
The gas is placed below the lower chain, and some of 
the particles are able to crossover up to the region above the upper chain. As 
time evolves, one reaches a steady state and we measure transmission by computing 
the number of particles in the upper region divided by the total number of 
particles.

Figure (\ref{fig4.ps}) shows the fraction of particles transmitted after many 
iterations, versus $\alpha$, for $k=5$ and $6$. On this range of $k$, as 
suggested by Fig.(\ref{fig2.ps}), there is no KAM curve in the phase space. In 
Fig.(\ref{fig4.ps}) we see an increase in particle transmission starting at 
the reconnection. The starting point of the increase can be evaluated 
analytically using Eq.(\ref{sepint}). For $k=5$ one obtains 
$\alpha=0.128$ in good agreement with the simulation. It is also observed that 
for $k=5$ the transmission reaches a plateau around $\alpha=0.15$; presumably 
at this value the reconnection is fully completed.

In Fig.(\ref{fig5.ps}) we finally display surface of sections of $20$ 
trajectories initially placed at $p=-2.0$ with uniform distribution along 
$\phi$. Fig.(\ref{fig5.ps}a) is made for a value of $\alpha$ prior to the 
reconnection, $\alpha = 0.10$, and Fig.(\ref{fig5.ps}b) with 
$\alpha$ past the reconnection threshold, $\alpha = 0.16$. The alterations 
in the diffusive pattern suggested by all the previous analysis can be 
seen in those figures as well - while in Fig.(\ref{fig5.ps}a) the 
particles remain mostly in the lower region, in Fig.(\ref{fig5.ps}b) 
particles can be easily transmitted across the barrier at $p \sim 0$.

To summarize, in this paper we have investigated the effect that reconnections 
involving unstable manifolds of deeply chaotic regimes can have on some 
macroscopically observable features like particle diffusion. We have 
used a nonmonotonic map to create reconnecting chains. Examining a 
particular family of fixed point for which $m=0$ in the notation of the text, 
we have seen that the topology of manifolds of unstable fixed points may 
have a similar behavior as in an integrable approximation. As we vary 
convenient parameters, manifolds of differents chains of islands can 
be clearly seen to undergo reconnection process even in deep chaotic 
regimes. This is somewhat unexpected, since if one lies in chaotic regimes 
where manifolds are already strongly interlaced, any reconnection would 
be expected of little influence in macroscopic features.  
\acknowledgments
FBR would like to acknowledge the hospitality at the Department 
of Electrical Engineering and Computer Science, UC Berkeley, where part of 
this work was developed. We deeply thank Prof. Allan Lichtenberg for his 
enlightening suggestions on the paper. GC has been supported by the Coordenadoria 
de Aperfei\c{c}oamento de Pessoal de N\'{\i}vel Superior (CAPES), Brazil and we 
receive additional support from Financiadora de Estudos e Projetos (FINEP) 
and Conselho Nacional de Desenvolvimento Cient\'{\i}fico e Tecnol\'ogico (CNPq), 
Brazil.


%
%
\centerline{FIGURE LEGENDS}

\begin{figure}[h]
\caption{Parameter space $(\alpha,k)$ for the period one $m=0$ resonance, and the 
relevant threshold curves.}
\label{fig2.ps}
\end{figure}
\begin{figure}[h]
\caption{The reconnection of stable and unstable manifolds in a deeply chaotic 
regime for which $k=5$.}
\label{fig3.ps}
\end{figure}
\begin{figure}[h]
\caption{Transmission fraction as a function of $\alpha$ in the deeply 
chaotic regimes $k=5$ and $k=6$.}
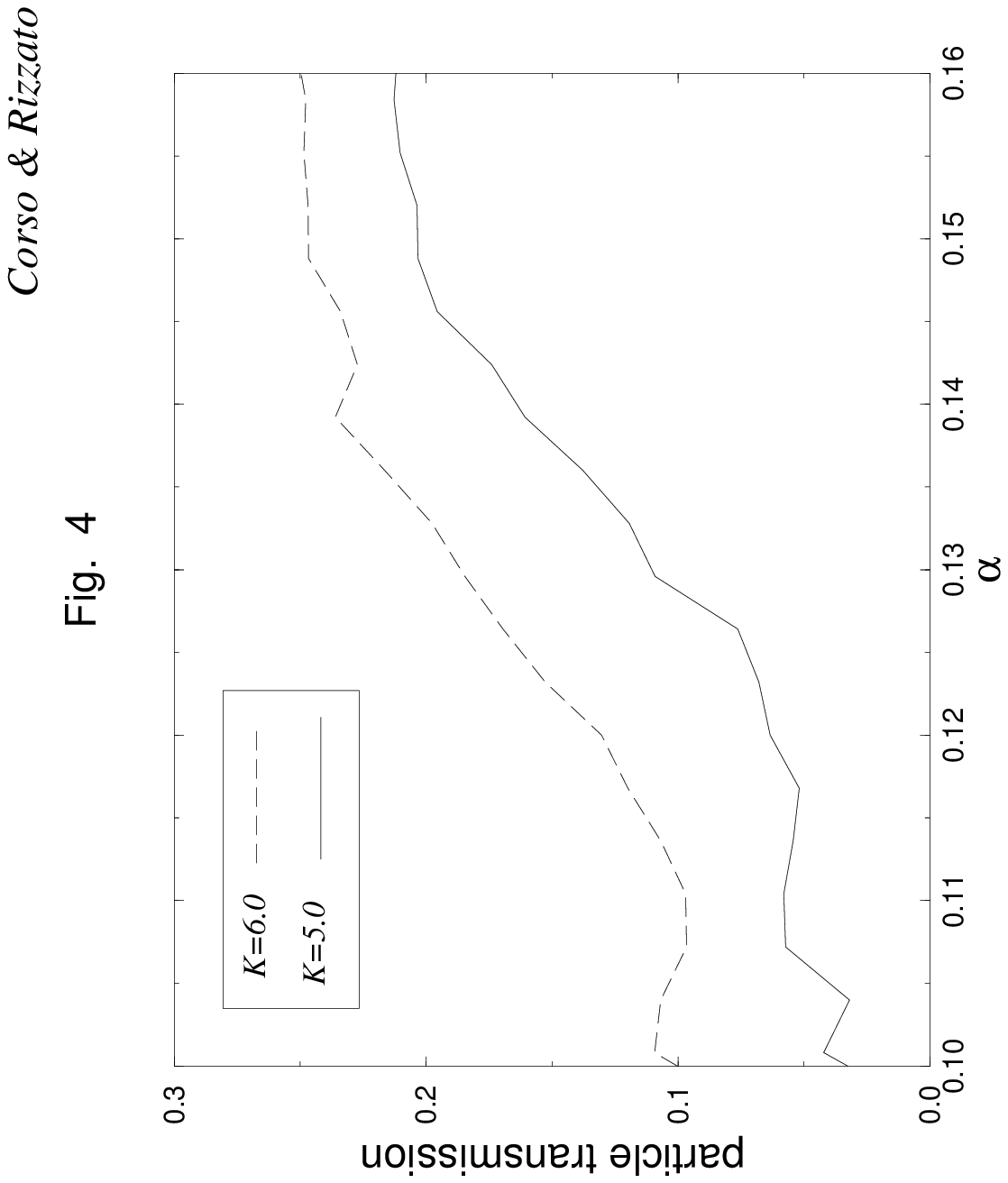
\label{fig4.ps}
\end{figure}
\begin{figure}[h]
\caption{Poincar\'e plot of the $k=5$ case for $\alpha=0.10$ in (a) and for 
$\alpha=0.16$ in (b).}
\label{fig5.ps}
\end{figure}
\end{document}